\documentstyle[twocolumn,aps,epsfig,floats]{revtex}

\begin{document}

\draft
\tolerance = 10000

\setcounter{topnumber}{1}
\renewcommand{\topfraction}{0.9}
\renewcommand{\textfraction}{0.1}
\renewcommand{\floatpagefraction}{0.9}

\twocolumn[\hsize\textwidth\columnwidth\hsize\csname
@twocolumnfalse\endcsname

\title{Noise-induced breakdown of coherent collective motion
in swarms}
\author{Alexander S. Mikhailov}
\address{Fritz-Haber-Institut der Max-Planck-Gesellschaft\\
Faradayweg 4-6, 14195 Berlin (Dahlem), Germany}
\author{Dami\'{a}n H. Zanette}
\address{Consejo Nacional de
Investigaciones Cient\'{\i}ficas y T\'ecnicas\\
Centro At\'omico Bariloche and Instituto Balseiro,
8400 Bariloche, Argentina}
\maketitle

\begin{abstract}
We  consider  swarms  formed  by  populations of self-propelled
particles  with  attractive  long-range  interactions.  These   swarms
represent multistable  dynamical systems  and can  be found  either in
coherent traveling states or in an incoherent oscillatory state  where
translational motion of the  entire swarm is absent.  Under increasing
the noise  intensity,  the  coherent  traveling  state  of  the swarms
is destroyed   and  an   abrupt  transition  to  the oscillatory state
takes place.
\end{abstract}


]

There  is  a  large  class  of  problems  where individual interacting
particles, that constitute a system, are capable of active motion  and
form collectively traveling populations. Self-propulsion of  particles
is   already   possible   in   simple   physical   systems  (see  e.g.
\cite{Dupeyrat,deGennes,Meinkohn,Kassner,Prost}) and  is widely  found
in biology where individual animals may group themselves into  swarms,
fish   schools,   bird   flocks   or   traveling   cell    populations
\cite{Alt,Alt-papers,Wissel,Gruler}.   The    role    of    individual
self-propelled ``particles'' can also be played by localized  patterns
(spots) in  reaction-diffusion systems.  A bifurcation  leading to the
onset  of  translational  motion  of  spots  has  been studied in an
activator-inhibitor system with global feedback \cite{Krischer} and in
three-component    reaction-diffusion    systems    \cite{Bode0,Bode}.
Interactions  between  individual   self-propelled  spots  have   been
determined from the  underlying reaction-diffusion equations  and used
to describe formation of bound states of such ``particles''
\cite{Bode1}.

Mathematical  modeling  of  collective  active  motion follows several
different directions. One approach is based on the notion of  discrete
stochastic automata\cite{Wissel,Reynolds,CV,Vicsek}.  Another approach
is formulated in terms of  continuous velocity and density fields  and
essentially  treates  a  swarm  as  an active fluid \cite{Toner} (such
hydrodynamical  equations  may  be  derived  by  averaging  from   the
respective automata models  \cite{Vicsek99}).  A  similar hydrodynamic
approach   is   also   used   in   the   theory   of   traffic   flows
\cite{Lighthill,Kerner}.   Alternatively,  one  can  specify dynamical
equations  of  motion  for  all  individual particles that explicitely
include interactions  between them  and/or action  of external  fields
\cite{Meinkohn,Alt-papers,Gruler,Bode1,Lutz}.  An  interesting
problem  related  to  statistical  mechanics  of  large populations of
self-propelled particles  is the  spontaneous development  of coherent
collective motion  in such  systems.   This problem  has recently been
discussed in the framework  of continuous hydrodynamical and  discrete
automata models, and  the properties of  the respective kinetic  phase
transition   were    numerically   and    analytically    investigated
\cite{Toner,Vicsek99}.   Both  in  one-  and  two-dimensional systems,
first- and second-order transitions have been found \cite{CV}.

In  the  present   paper  we  consider   a  population  of   identical
self-propelled  particles   near  a   transition  between   disordered
oscillating motion and coherent  translational motion.  The  particles
interact via an isotropic attractive binary potential and are  subject
to the  action of  noises. This  globally coupled  population forms  a
cloud (the swarm) in the considered one-dimensional space.  The  swarm
can be found in different  states.  Coherent compact traveling  states
are  characterized  by  a  narrow  distribution of velocities around a
certain mean  drift velocity,  directed either  to the  left or to the
right. Another possible  state of this  population corresponds to  the
absence  of  coherent  translational  motion,  with noisy oscillations
around a  certain mean  position in  space, determined  by the initial
conditions.

The coherent traveling states  exist only for sufficiently  weak noise
and,  as  the  noise  intensity  increases,  the  swarm  undergoes   a
transition  to  the  incoherent  oscillatory  state.  We find that the
breakdown of coherent collective motion  in this system is abrupt  and
characterized by a strong hysteresis. Thus, the globally coupled swarm
represents a multistable system that may be found in different  states
depending  on  the  initial  conditions.   This  behavior, revealed by
numerical simulations, is well reproduced by an approximate analytical
theory and may represent a typical property of swarms with  long-range
interactions.

To formulate  the model,  we note  that if  a system  is close  to the
onset  of   active  motion   and  this   instability  is   soft,  i.e.
characterized by  a supercritical  bifurcation, the  motion with small
velocity $V$ can generally be described by equation
\begin{equation}
\dot {V}=\alpha V-\beta V^{3},  \label{GL}
\end{equation}
with real coefficients $\alpha $ and $\beta >0$. This equation may  be
viewed as a  normal form of  the supercritical bifurcation  leading to
translational  motion.  Such  bifurcations  are  possible  in   simple
physico-chemical  systems  \cite{Meinkohn}.  They  are  also known for
localized spot patterns in reaction-diffusion models and correspond to
the onset of their translational motion \cite{Krischer,Bode}.

According  to  Eq.  (\ref{GL}),  the  velocity  $V$  is zero below the
bifurcation point  (i.e. for  $\alpha <0$).  Above this  point, active
motion  with   $V=\pm  \sqrt{\alpha   /\beta  }$   is   asymptotically
established. The direction of  this motion for an  individual particle
remains arbitrary and is  determined by initial conditions.  Rescaling
time and introducing the new velocity variable $u=V\sqrt{\beta /\alpha
}$, Eq. (\ref{GL}) can be written as
\begin{equation}
\dot {u}=u-u^{3}  \label{GL1}
\end{equation}
When a  population of  identical self-moving  particles is considered,
the velocity $u_{i}=\dot  {x}_{i}$ of each  particle $i$ will  satisfy
this dynamical equation.

Interactions between  individuals may  generally depend  on both their
relative positions and  velocities. In this  paper we assume  that the
interactions  are  pairwise  and  described by forces $f(x_{i}-x_{j})$
that depend only on the difference of coordinates of two particles $i$
and $j$. We shall further assume that the interactions are  attractive
and  depend  linearly  on  the  distance  between  the particles, i.e.
$f(x_{i}-x_{j})\propto (x_{i}-x_{j})$.   These attractive  forces  are
supposed  to  model  the  interaction  within  the  size ranges of the
dynamical states considered below,  where the population forms  clouds
of either oscillating or  translational motion. The interaction  could
be extended to  larger distances in  order to represent, for instance,
vanishing forces  at infinity  \cite{Shimo}. Additionally,  the system
may include noise that will  be modelled by independent random  forces
$\xi  _{i}(t)$  acting  on  individual  particles.  Noise prevents the
collapse  of   the  population,   so  that   short  range    repulsion
\cite{CV,Shimo} can here be ignored.

Under these conditions, the dynamical equations for a set of $N$
identical self-moving particles with coordinates $x_{i}(t)$ are
\begin{equation}
\ddot{x}_{i}+(\dot{x}_{i}^{2}-1)\dot{x}_{i}+\frac{a}{N}\sum_{j=1}^{N}
\left( x_{i}-x_{j}\right) =\xi _{i}(t),
\label{mod}
\end{equation}
for $i=1,\dots N$.
The coefficient  $a$ characterizes  the intensity  of interactions and
can be viewed as the parameter, specifying the strength of coupling in
the  population.  Equations  (\ref{mod})  constitute  the  basic model
investigated in  this paper.  We shall  assume that  $\xi _{i}(t)$ are
independent  white  noises  of  intensity  $S$,  so  that $\langle \xi
_{i}(t)\xi _{j}(t')\rangle  =2S\delta _{ij}\delta  (t-t')$. Note  that
Eqs.  (\ref{mod})  are   invariant  with  respect   to  an   arbitrary
translation in the coordinate space.

The  model  (\ref{mod})  can  behave  as  a system of globally coupled
limit-cycle    oscillators    (cf.    \cite{Kuramoto,Nakagawa}).
Introducing the average coordinate $\overline{x}(t)$ of the swarm,
\begin{equation}
\overline{x}(t)=\frac{1}{N}\sum_{j=1}^{N}x_{j}(t),  \label{aver}
\end{equation}
Eqs. (\ref{mod}) in absence of noise read
\begin{equation}
\ddot{x}_{i}+(\dot{x}_{i}^{2}-1)\dot{x}_{i}+a\left(
x_{i}-\overline{x} \right) =0\ \ \ \ \ (i=1,\dots ,N).  \label{mod1}
\end{equation}
Thus, if the swarm does not move as a whole, i.e. $\overline{x} (t)  =
\mbox{constant}$,  the particles perform persistent oscillations.   In
this  state  the  phases  of   individual
oscillations are random. Note that the spatial location $\overline{x}$
of an oscillating swarm is arbitrary.

In addition to the  random oscillatory state, the  system (\ref{mod1})
has  two  coherent  collapsed  states  where  the  coordinates  of all
particles are identical, i.e. $x_{i}= \overline{x}$ for any $i$. These
states correspond to uniform translational motion of the entire  swarm
with  the  velocity  $u=\pm  1$.  A  simple  analysis  shows  that the
oscillatory  state  and  both  coherent  traveling states are linearly
stable  for  any  positive  parameter  $a$.  The  final  state  of the
population  is  determined  by  the  initial conditions. Our numerical
simulations  show  that,  if   the  average  velocity    $\overline{u}
=N^{-1}\sum_{i}u_{i}$  is  initially  close  to  zero, the oscillatory
standing state  is asymptotically  reached. If,  however, this initial
average velocity is  large enough, one  of the two  coherent traveling
states will be approached.

\begin{figure}
\begin{center}
\psfig{figure=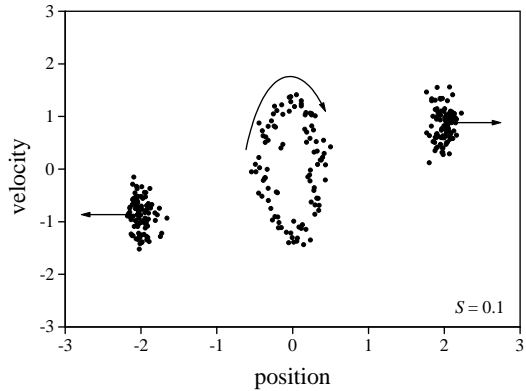,width=\columnwidth}
\end{center}
\caption{Three  snapshots  of  100-particle  systems  with  $a=10$ and
$S=0.1$,  in  different  dynamical   regimes.  The  central   ensemble
corresponds to disordered oscillations along a noisy limit cycle.  The
other two ensembles  stand for coherent  clouds with opposite  average
velocities.}
\end{figure}

Since the particles either  converge to coherent motion  with constant
velocity  or  to  disordered  oscillations  with no average drift, the
ensemble  can  be  thought  of  as  a  {\it  multistable  system} with
qualitatively different  attractors. In  the following,  we focus  the
attention on how these attractors respond to the effect of noise. With
this  aim,  we  study  Eq.  (\ref{mod})  numerically.  Integration  is
performed by means of a standard Euler scheme with a time step $\Delta
t=10^{-3}$ to $10^{-2}$. Most calculations correspond to ensembles  of
$100$ particles,  with the  coupling intensity  ranging from  $a=1$ to
$100$. Larger  values of  $a$ require  smaller values  of $ \Delta t$.
Noise is introduced by generating, at each time step, a random  number
$\xi  $  with  uniform  distribution  in  the interval $(-\xi _{0},\xi
_{0})$. This choice corresponds to having $S=\xi _{0}^{2}/6\Delta  t$.
In practice,  $\xi _{0}$  is calculated  for each  given value of $S$.
Initial conditions are selected at random, distributing the  particles
around $x=0$ and $u=0$ or $1$ with a dispersion of the order of  $0.5$
in both variables.  From each initial condition the system is left  to
evolve  in  the  absence  of  noise  until  it  reaches  the  state of
disordered oscillations or coherent motion. Then, at $t=30$, noise  is
switched on. Typical calculations extend up to $t\approx 1000$.

For  small   noise  intensities,   $S\lesssim  0.1$,   the  stochastic
perturbations to the trajectories preserve the characteristic features
of  the  collective  dynamics  observed  in  the absence of noise. The
completely collapsed  state of  the noiseless  case transforms  into a
cloud of particles which still  moves coherently at a given  velocity.
Oscillatory orbits, meanwhile, proceed now along a noisy limit  cycle.
Figure 1 shows three snapshots  of a system of 100  particles
with  $a=10$,  subject  to  noise  with  $S=0.1$.   They  started from
different initial conditions, as described above.  The arrows indicate
the overall motion of each swarm.

Within coherent  clouds, each  particle performs  an oscillatory noisy
motion  which  is  superimposed  to  the  collective  translation. The
distribution  of  particles  inside  the  clouds  has  a  well defined
profile, shown in Fig. 2 for  some values of $a$ in the  case
of positive velocity.  The normalized distribution $\rho (y)$ is there
plotted  as  a  function  of  the  coordinate  relative to the average
position, $y_{i}=x_{i}-\bar{x}$. For decreasing $a$, the  distribution
becomes broader and more asymmetric, with an accumulation of particles
at the front of the cloud.

\begin{figure}
\begin{center}
\psfig{figure=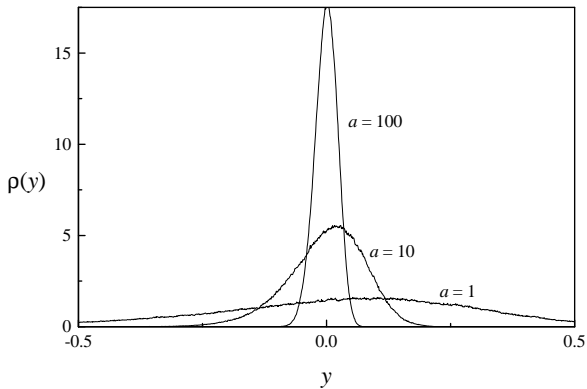,width=\columnwidth}
\end{center}
\caption{Normalized profiles  of coherent  clouds as  functions of the
deviation from the average position,  for different values of $a$  and
$S=0.1$ in a 100-particle ensemble. }
\end{figure}

The coherent  traveling states  of the  population cease  to exist  at
sufficiently high noise intensities and the swarm undergoes an  abrupt
transition  to  its  random  oscillatory  state,  characterized by the
absence of the translational motion. This breakdown of coherent  swarm
motion is illustrated in Fig.   3.  We see that if  the noise
is relatively weak (Fig. 3a), switching it on at $t=30$  only
produces a slight decrease of  the velocity of the coherent  cloud, so
that the  average velocity  $\overline{u} (t)  $ exhibits fluctuations
around a constant mean value  $\overline{u}<1$.  If however the  noise
intensity exceeds a certain threshold, the effect of introducing noise
is qualitatively different (Fig. 3b).  Within a certain  time
interval  after  the  introduction  of  noise,  the swarm continues to
travel at  a somewhat  reduced, strongly  fluctuating average velocity
$\overline{u}(t)$.  Then,  it suddenly starts  to decelerate and  soon
reaches a  steady state  where the  average velocity $\overline{u}(t)$
fluctuates near zero. Inspection  of the distribution of  particles in
the ensemble shows that in this state the system has been attracted to
the noisy limit  cycle mentioned above.   We conclude that  the system
undergoes  a  {\it  noise-induced  transition}  from  a  condition  of
multistability with two kinds of attractors to a situation where  only
one of  them exists.   The coherent  clouds observed  for small  noise
intensities are no  longer possible for  $S>S_{c}$, and the  system is
necessarily led to the state of noisy, disordered oscillations.

\begin{figure}
\begin{center}
\psfig{figure=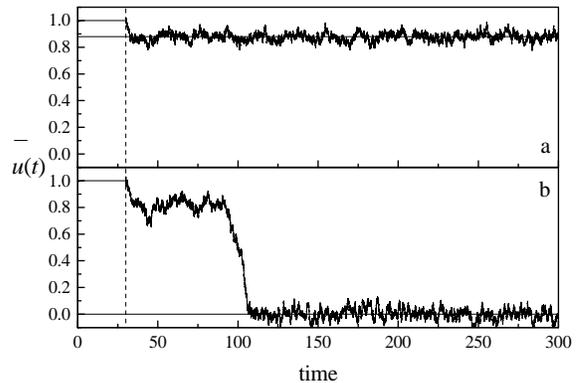,width=\columnwidth}
\end{center}
\caption{Average  velocity  of  100-particle  coherent  clouds  in two
realizations at  (a) $S=0.10$  and (b)  $0.12$, with  $a=10$. Noise is
switched on at $t=30$ (dashed line). The horizontal lines indicate the
asymptotic mean values of $u(t)$. }
\end{figure}

Figure  4  displays  the  dependence  of  the  mean  velocity
$\overline{u}$ of the traveling swarm  on the noise intensity $S$  for
three different values  of the coupling  coefficient $a$. We  see that
the mean  velocity monotonously  decreases with  the noise  intensity,
until a certain critical noise  intensity is reached and the  coherent
swarm motion  becomes impossible.  The mean  velocity at  the critical
point  is  still  relatively  large,  $\overline{u}  \approx 0.8.$ The
critical noise intensity $S_{c}$  becomes lower for smaller  values of
$a$. Note that the behavior of the swarm is characterized by a  strong
hysteresis.   If the  breakdown of  the coherent  motion has occurred,
subsequently decreasing the noise  intensity leaves the system  in the
oscillatory state with zero mean velocity, down to $S=0$.

\begin{figure}
\begin{center}
\psfig{figure=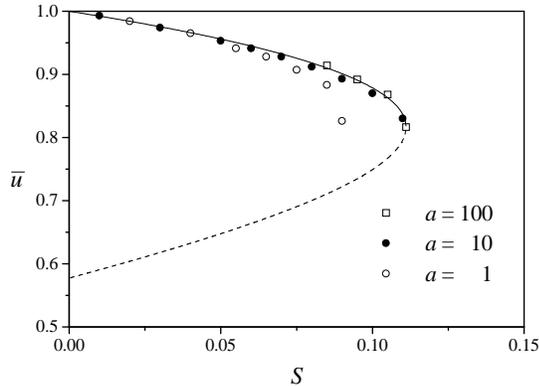,width=\columnwidth}
\end{center}
\caption{The asymptotic mean velocity of 100-particle coherent  clouds
as a function of $S$, for different values of $a$. Dots correspond  to
numerical measurements and lines  stand for the analytical  result Eq.
(\ref {uS}). }
\end{figure}

An interesting property of the considered noise-induced transition  is
the divergence of the waiting time at the critical point. The  waiting
time $ T_{0} $  is defined as the  time at which the  average velocity
$\overline{u} (t) $ of the  cloud first reaches zero (we  measure this
time starting from the moment  $t=30$ when the noise is  switched on).
Figure 5  shows the  waiting time  $T_{0}$ as  a function  of
$S-S_{c}$ in  a log-log  plot. We  see that  for very  small values of
$S-S_{c}$, this  time decreases  following a  power law, $T_{0}\propto
(S-S_{c})^{-\gamma  }$,  with  $\gamma  \approx  1.33$. Then, at about
$S-S_{c}=0.03$,  the  behavior  changes  to  a  power law with $\gamma
\approx 0.52$.   Straight dashed lines  with slopes $-4/3$  and $-1/2$
have been plotted for reference.

\begin{figure}
\begin{center}
\psfig{figure=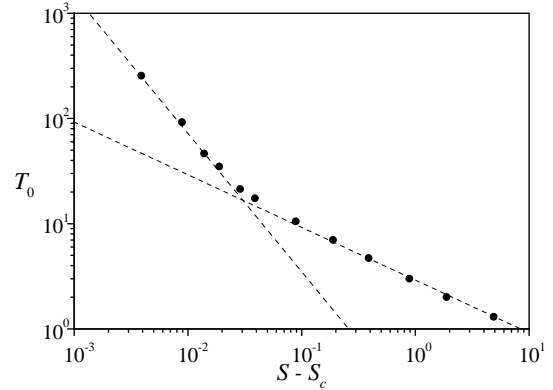,width=\columnwidth}
\end{center}
\caption{Waiting  time  $T_0$,  at  which  the  average  velocity   of
initially coherent clouds vanishes for  the first time, as a  function
of $ S-S_c$ for $a=10$ in a 100-particle system. The dashed lines have
slopes $ -1/2$ and $-4/3$. }
\end{figure}

The  observed  noise-induced  transition  between  coherent clouds and
disordered oscillations  of the  swarm can  be explained  by a  simple
approximate  analytical  approach.  By  summing  all Eqs. (\ref{mod1})
for different particles  $i$ and taking  into account that  the noises
acting  on  individual  particles  are  not  correlated,  an evolution
equation for the average swarm velocity $\overline{u}(t)$ is obtained:
\begin{equation}
\dot{\overline{u}}+\frac{1}{N}\sum_{i=1}^{N}\dot {x}_{i}^{3}-
\overline{u}=0 . \label{eq1}
\end{equation}
Let us  introduce for  each particle  its deviation  $y_{i} =  x_{i} -
\bar{x}$ from the average position of the swarm. Then we can write
\begin{equation}
\frac{1}{N}\sum_{i=1}^{N}\dot {x}_{i}^{3}=\overline{u}^{3}+3\sigma
\overline{u}+\frac{1}{N}\sum_{i=1}^{N}\dot {y}_{i}^{3} , \label{eq2}
\end{equation}
where $\sigma  =N^{-1}\sum_{i}\dot{y}_{i}^{2}$ is  the average  square
dispersion of the swarm. The last  cubic term in this equation can  be
neglected if the distribution of  particles in the traveling cloud  is
symmetric.  As  we   have  seen  from   numerical  simulations   (Fig.
2),  this  is  indeed  a  good approximation for sufficiently
large values of the coupling constant $a$. Within this  approximation,
Eq. (\ref{eq1}) takes the form
\begin{equation}
\dot{\overline{u}}+(\overline{u}^{2}-1)\overline{u}+3\sigma
\overline{u}=0.  \label{u}
\end{equation}

On the  other hand,  deviations of  particles from  the center  of the
swarm obey the stochastic differential equation
\begin{eqnarray}
\ddot{y}_{i}+(3\overline{u}^{2}-1)\dot{y}_{i}&+&ay_{i}+3\overline{u}
\left(\dot {y}_{i}^{2}-\sigma \right) \nonumber \\
&+&\left( \dot {y}_{i}^{3}-\frac{
1}{N}\sum_{i=1}^{N}\dot {y}_{i}^{3}\right) =\xi _{i}(t).  \label{eq3}
\end{eqnarray}
Assuming that the deviations of $\dot {y}_{i}$ are relatively small
and linearizing this equation, we obtain
\begin{equation}
\ddot{y}_{i}+(3\overline{u}^{2}-1)\dot{y}_{i}+ay_{i}=\xi _{i}(t).
\label{y}
\end{equation}
In  this  approximation  the  deviations  for  different particles $i$
represent statistically independent  random processes. This  allows us
to replace  the ensemble  average in  the dispersion  $\sigma $ by the
statistical average taken over independent random realizations of such
processes, defined by Eq. (\ref{y}).

Hence,  we  have  derived  a  closed  set  of  equations (\ref{u}) and
(\ref{y})  that  approximately   describe  the  swarm.   We  want   to
investigate steady statistical states  of this system. The  stationary
solutions to  Eq. (\ref{u})  are $\overline{u}=\pm  \sqrt{1-3\sigma }$
and $u=0$. The latter solution corresponds to the resting swarm.

Examining Eq. (\ref{y}), we note that it describes damped oscillations
only if  $3\overline{u}^{2}-1>0$, i.e.  only if  mean velocity  of the
swarm  is  sufficiently  large.  Under  this condition, the stationary
probability distribution for $y_{i}$ is readily found and the  average
square dispersion of velocities is obtained as
\begin{equation}
\sigma =\frac{S}{3\overline{u}^{2}-1}.  \label{sigma}
\end{equation}

The algebraic equations for $\overline{u}$ and $\sigma$ can be  solved,
yielding  the  statistical  dispersion  of  particles in the traveling
swarm,
\begin{equation}
\sigma _{1,2}=\frac{1}{9}\left( 1\pm \sqrt{1-9S}\right) ,  \label{sig}
\end{equation}
and its mean velocity
\begin{equation}
\overline{u}^2_{1,2}=\frac{1}{3}\left( 2\pm \sqrt{1-9S}\right)
.  \label{uS} \end{equation}

Thus, the traveling state solutions disappear when the critical  noise
intensity $S_{c}=1/9=0.11\dots  $ is  reached. At  this critical point
the mean swarm velocity is $\overline{u}_{c} = \sqrt{2/3} =  0.82\dots
$  and  the  mean  dispersion  of  particles  in  the cloud is $\sigma
_{c}=1/9=0.11\dots$.

Below the breakdown threshold (for $S<S_{c}$), solution (\ref{uS}) has
two branches shown by solid  and dashed lines in Fig. 4. The
lower branch  is apparently  unstable, since  it approaches  the value
$\overline{u}=1/\sqrt{3}=0.58\dots$ at $S=0$,  i.e. in absence  of the
noise. A special property of the derived solution is that it does  not
depend on the parameter $a$.

Comparing the theoretical  prediction with the  numerically determined
values  of  the  mean  swarm  velocity,  that are also plotted in Fig.
4, we can see that this approximation provides good estimates
of  the  swarm  velocity  and  the  critical  noise intensity when the
parameter $a$ is relatively high ($a=100$ and $a=10$). At small values
of $a$, the deviations  from the numerical results  become significant
near the breakdown threshold. This  can be understood if we  take into
account that, according to Fig. 2, the distribution of  particles
in a  traveling swarm  shows significant  asymmetry for  such a  small
value of $a$ and therefore our approximations are not valid.

For a standing swarm ($\overline{u}=0$), the deviations  $y_{i}=x_{i}-
\overline{x}$ obey in the  limit $N\rightarrow \infty $  the nonlinear
stochastic differential equation
\begin{equation}
\ddot{y}_{i}+(\dot{y}_{i}^{2}-1)\dot{y}_{i}+ay_{i}=\xi _{i}(t),
\label{y0}
\end{equation}
which  is  similar  to  the  Van  der Pol equation \cite{VdP}. In this
state,  therefore,  the  particles  in  the  swarm  perform   periodic
limit-cycle oscillations  with a  random distribution  of phases. This
state exists for  any noise intensity  $S$ and is  approached when the
noise-induced  breakdown  of  the  coherent  motion  takes  place   at
$S=S_{c}$.

Thus,  we  have  find  in  this  paper  that  a  swarm of interacting,
actively moving particles may  show bistable behaviour, i.e.  be found
either in a  coherent state, traveling  at a fixed  velocity, or in  a
rest state where the translational motion is absent and the individual
particles perform  oscillations around  the center  of the  swarm. The
bistability persists in the presence of noise if its intensity remains
relatively  low.  Increasing  the  noise  intensity  leads to a sudden
breakdown of  the coherent  traveling motion  and a  transition to the
resting oscillatory state occurs. This behavior is different from  the
second-order  phase  transitions  to  coherent collective motion, that
were found in the previously studied models \cite{Toner,Vicsek99}.  We
conjecture that  the difference  is related  to the  fact that  in our
model the  interactions between  self-propelled particles  have a long
range and extend over the entire swarm. It would be interesting to see
how this behavior is modified when other interaction laws and  systems
with higher dimensionality are  considered. Finally, we   remark that,
when  formulated  in  terms  of  dynamical  equations  for  individual
interacting self-propelled  particles, the  problem shows  significant
similarities  to  synchronization  and  condensation in populations of
globally coupled oscillators (see e.g. \cite{Kuramoto,zanette}).
The  significant  new  aspect  is  that  collapsed  synchronous states
correspond here to translational motion of the entire population.

The authors acknowledge financial support from Fundaci\'on Antorchas
(Argentina).

\end{document}